\documentclass[final,5p,twocolumn]{elsarticle}

\pdfoutput=1

\usepackage{amsfonts}
\usepackage{amssymb}
\usepackage{amsmath}
\usepackage{xcolor,colortbl}
\usepackage[breaklinks,bookmarks=false,pdfstartview={XYZ null null 1.00}]{hyperref}
\usepackage{tikz}\usetikzlibrary{arrows,shapes.geometric,positioning,shapes}
\usepackage{balance}

\biboptions{numbers,sort&compress}
\journal{Elsevier Big Data Research. Published in vol.~2, no.~3, pp.~94-101, Sep.~2015}

\begin{document}

\begin{frontmatter}

\title{Big Data Analytics for Dynamic Energy Management in Smart Grids}

\author[Aristotle]{Panagiotis D. Diamantoulakis\corref{mycorrespondingauthor}}
\ead{padiaman@auth.gr}
\author[Aristotle]{Vasileios M. Kapinas}
\ead{kapinas@auth.gr}
\author[Khalifa,Aristotle]{George K. Karagiannidis}
\ead{geokarag@auth.gr}

\address[Khalifa]{Department of Electrical and Computer Engineering, Khalifa University, PO Box 127788, Abu Dhabi, United Arab Emirates}
\address[Aristotle]{Department of Electrical and Computer Engineering, Aristotle University of Thessaloniki, GR-54124 Thessaloniki, Greece}
\cortext[mycorrespondingauthor]{Corresponding author}




\begin{abstract}
The smart electricity grid enables a two-way flow of power and data between suppliers and consumers in order to facilitate the power flow optimization in terms of economic efficiency, reliability and sustainability. This infrastructure permits the consumers and the micro-energy producers to take a more active role in the electricity market and the dynamic energy management (DEM). The most important challenge in a smart grid (SG) is how to take advantage of the users' participation in order to reduce the cost of power. However, effective DEM depends critically on load and renewable production forecasting. This calls for intelligent methods and solutions for the real-time exploitation of the large volumes of data generated by a vast amount of smart meters. Hence, robust data analytics, high performance computing, efficient data network management, and cloud computing techniques are critical towards the optimized operation of SGs. This research aims to highlight the big data issues and challenges faced by the DEM employed in SG networks. It also provides a brief description of the most commonly used data processing methods in the literature, and proposes a promising direction for future research in the field.
\end{abstract}

\begin{keyword}
Big data \sep Smart grids \sep Dynamic energy management \sep Predictive analytics \sep Artificial intelligence \sep high performance computing.
\end{keyword}

\end{frontmatter}

\section{Introduction}\label{sec:intro}
A smart grid (SG) is the next-generation power system able to manage electricity demand in a sustainable, reliable and economic manner, by employing advanced digital information and communication technologies. This new platform aims to achieve steady availability of power, energy sustainability, environmental protection, prevention of large-scale failures, as well as optimized operational expenses (OPEX) of power production and distribution, and reduced future capital expenses (CAPEX) for thermal generators and transmission networks~\cite{Bush}. The upcoming technology in the framework of SG facilitates the development and efficient interactive utilization of millions of alternative distributed energy resources (DER) and electric vehicles~\cite{Bush, Goncalves, Lopez}. To this end, each consumer location has to be equipped with a smart meter for monitoring and measuring the bi-directional flow of power and data, while supervisory control and data acquisition (SCADA) systems are needed to control the grid operation.

\begin{figure}
\centering%
\centering\includegraphics[width=1\linewidth,trim=0 0 0 0,clip=false]{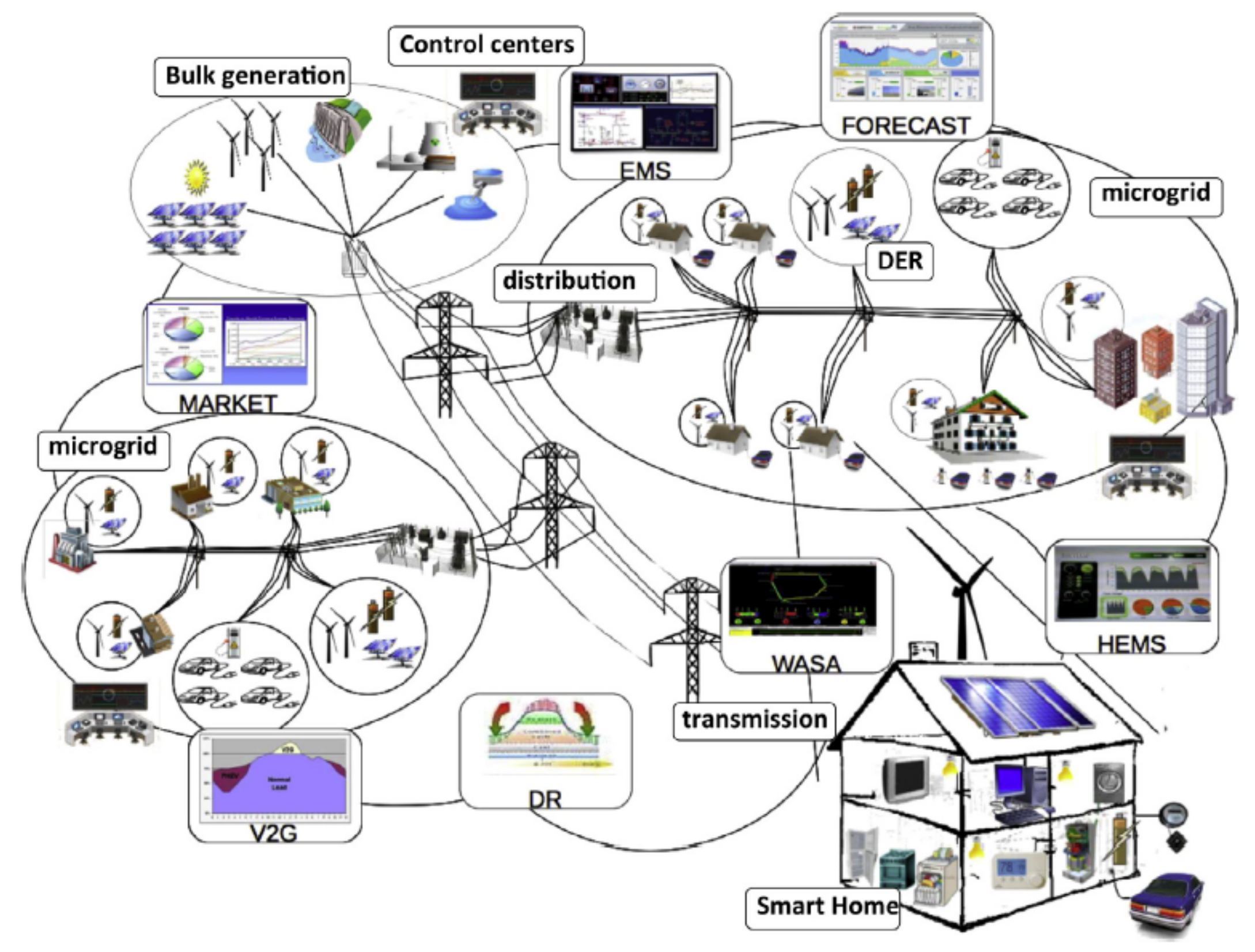}
\caption{Smart Grid overview~\cite{Ancillotti}.}
\label{systemmodel}
\end{figure}

While dynamic energy management (DEM) in conventional electricity grids is a well-investigated topic, this is not the case for SGs. This is due to its much more complicated nature, since complex decision-making processes are required by the control centers~\cite{Mallik,Balac}. Energy management systems (EMSs) in SGs include i) real-time wide-area situational awareness (WASA) of grid status through advanced metering and monitoring systems, ii) consumers' participation through home EMSs (HEMS), demand response (DR) algorithms, and vehicle-to-grid (V2G) technology, and iii) supervisory control through computer-based systems~\cite{Ancillotti}. A typical overview of the SG and the included systems and technologies is given in Fig.~\ref{systemmodel}. The quality and reliability of the data collected is a key factor for the optimized operation of the SG, thus rendering data mining and predictive analytics tools essential for the effective management and utilization of the available sensor data~\cite{Zhong}. This is because effective DEM relies dramatically on short-term power supply and consumption forecasting, which handles prediction horizons from one hour up to one week~\cite{Mirowski}. Additionally, the sensor data contains important correlations, trends, and patterns that need to be exploited for the optimization of the energy consumption and the DR, among others~\cite{Mallik}. Most of the research related to data mining in SGs deal with predictive analytics and load classification (LC), which are necessary for the load forecasting, bad data correction, determination of the optimal energy resources scheduling, and setting of the power prices~\cite{Zhou, Vale}. The efficient processing of the produced vast amount of data requires increased data storage and computing resources, which imply the need for high performance computing (HPC) techniques.

This work differs from other related surveys in the literature, such as~\cite{Ukil, Simmhan, Leeds, Stimmel, Nguyen,IBM}, in being the first meta-analytic review on efficient SG data processing with focus on DEM. Besides, it gives useful insights into technologies and methods from the area of big data analytics (BDA) that have to be further explored into the framework of DEM, demand forecasting, and dynamic pricing. Our investigations reveal that there is free space for research in the following topics:
\begin{itemize}
\item Design and development of algorithms that can accurately extract the load patterns from large-scale datasets.
\item Design of machine learning (ML)-based algorithms with improved forecasting performance, low memory requirements, and scalable architecture.
\item Development of novel data-aware resource management systems that can provide powerful data processing in distributed computing systems and clusters for real-time processing.
\end{itemize}
It is also highlighted that scalability and flexibility, achieved through the construction of robust algorithms and fast provisioning of HPC resources, can enable the efficient processing of the large data volumes involved in DEM and short-term power demand/supply forecasting.

The rest of the paper is organized as follows. Section~\ref{sec:energymanagement} gives insight on the reasons why conventional data processing techniques are not appropriate for DEM in SGs. Section~\ref{sec:dataminingsg} focuses on smart meter data stream mining and presents the most commonly used methods. Section~\ref{sec:HPC} is dedicated to the appropriate HPC techniques. Section~\ref{sec:futuredirs} provides promising future research directions, while Section~\ref{sec:conclusion} concludes the paper.

\section{Dynamic Energy Management in SGs: A Big Data Issue}\label{sec:energymanagement}
DEM requires power flow optimization, system monitoring, real-time operation, and production planning~\cite{Manfren}.
In more detail, DEM in a SG is a complicated, multi-variable procedure, since the latter enables an interconnected power distribution network by allowing a two-way flow of both power and data. This is in contrast to the traditional power grid, in which the electricity is generated at a central source and then distributed to consumers. Thanks to the bi-directional flow of information and power between suppliers and consumers, the grids become more adaptive to the increased penetration of DER, encouraging also users' participation in energy savings and cooperation through the DR mechanism~\cite{Samadi, Mohsenian1, Vale}.

DR can be applied to both residential (e.g., cooling, heating, electric vehicles (EVs) charging, etc.) and industrial loads and includes three different concepts; i) energy consumption reduction, ii) energy consumption (or production) shifting to periods of low (or high) demand, and iii) efficient utilization of storage systems~\cite{Siano}. It should be noticed here that plug-in EVs can be considered as storage devices, while the careful scheduling of their charging and discharging can benefit both their owners and the utilities. Obviously, this further increases the parameters that the DEM algorithms have to take into account, such as the EVs charging profiles. Consequently, the associated complexity is also increased, creating at the same time storage capacity prediction problems~\cite{Wang1}. Thus, a crucial issue in SGs is how to manage DR in order to reduce peak electricity load, utilizing at the same time renewable energies and storage systems more efficiently. Finally, effectiveness of DR algorithms depends critically on demand, price, load, and renewable energy forecasting, which highlights the need for sophisticated signal processing techniques~\cite{Chan}.

The electricity demand and renewable production in the SG environment is affected by several factors, including weather conditions, micro-climatic variations, time of day, random disturbances, electricity prices, DR, renewable energy sources, storage cells, micro-grids, and the development of EVs~\cite{Mohsenian2, Carpinelli, Saber, Avila}. High forecasting accuracy accommodates the generation and transmission planning, i.e., deciding which power plants to operate and how much power should be generated by them at a specific time-period, with the aim to reduce the operating cost and increase the reliability~\cite{Balantrapu}. It also enables the utilities to successively estimate the electricity cost and correctly set the electricity prices, capturing the interdependency between the energy demand and the prices~\cite{Motamedi}. A typical example of this interdependency is load-synchronization, where a large portion of load is shifted from hours of high prices to hours of low prices, without significantly reducing the peak-to-average ratio~\cite{Mohsenian2}. Moreover, insufficient monitoring and control of the power flow can increase the possibility of failure (e.g., due to load synchronization, overloading, congestion, etc.). The power grid, which is consisted of multiple components such as relays, switches, transformers, and substations, must be carefully monitored. Therefore, the SG requires intelligent real-time monitoring techniques in order to be capable of detecting abnormal events, finding their location and causes, and most importantly predicting and eliminating faults before they happen. This self-healing behaviour, renders the power grid a real ``immune system", which is one of the most important characteristics of a SG framework, targeting uninterrupted power supply~\cite{Pitt, Jia}.

In order to deal with the high level of uncertainties in DEM, the extreme size of data, and the need for real-time learning/decision making, the SG demands advanced data analytic techniques, big data management, and powerful monitoring techniques~\cite{Aung, Mack}. Since BDA is one of the major driving forces behind a SG, various techniques such as artificial intelligence, distributed and HPC, simulation and modeling, data network management, database management, data warehousing, and data analytics are to be used to guarantee smooth running of SGs. The main challenges of Big Data approaches in SGs is the selection, deployment, monitoring, and analysis of aggregated data in real-time~\cite{Baek}. The BD's role in SG collective awareness, the self-organization capability of SGs, and the service interruptions limitation are thoroughly discussed in~\cite{Pitt}. Specifically, it is explained that the reliability of the electricity grid could be enhanced if the users were aware of the effects of their personal energy use on the total consumption and overloading.

Considering the above, BDA can provide efficient solutions in specific problems related to data processing in SGs as described in the sequel and briefly summarized in Fig.~\ref{roadmap}, which illustrates the roadmap to the DEM in SGs.

\definecolor{blue1}{gray}{0.8}
\definecolor{blue2}{gray}{0.9}

\begin{figure*}
\centering%
\resizebox{0.9\linewidth}{!}{%
\begin{tabular}{p{0.5\linewidth}p{0.5\linewidth}}
\hline

\textbf{Concern} & \textbf{Solution}\\
\hline
\vspace{1em} DR is huge challenge in practise since each user has a different reaction to the real-time prices and stochastic parameters such as weather, etc.~\cite{Vale}. Also, many inserted factors are interdependent~\cite{Motamedi}.\vspace{1em} & \vspace{1em} Before optimizing the power flow and setting the prices, LC can be used in order to categorize various load patterns into a specific number of groups and each group can be characterized by each characteristic load pattern~\cite{Vale}. Predictive analytics and ML techniques can facilitate the real-time decision making~\cite{Mack}.\vspace{1em}\\

\rowcolor{blue2}
\vspace{1em} The available communication resources, e.g., bandwidth are not enough for the data acquisition in a centralized manner~\cite{Dieb}.\vspace{1em} & \vspace{1em} Distributed data mining and dimensionality reduction reduce the required resources~\cite{Dieb, Mallik, Bhaduri}.\vspace{1em}\\

\vspace{1em} Traditional computer techniques cannot handle fast data processing, which is required for real-time monitoring, dynamic energy management and power flow optimization~\cite{Green}.\vspace{1em} & \vspace{1em} Efficient high performance computing techniques, such as virtualization and in-memory computing, can importantly reduce the computing time~\cite{Green, Ali}.\vspace{1em}\\

\rowcolor{blue2}
\vspace{1em} Increased data storage and computing resources are required in order to meet the needs of processing the huge amount of data involved in DEM, which could possibly lead to increased OPEX and CAPEX~\cite{Ali}.\vspace{1em} & \vspace{1em} Cloud computing: pay-per-use~\cite{Rusitschka}.\vspace{1em}\\

\vspace{1em} Security~\cite{Yigit, raey}\vspace{1em} & \vspace{1em} Data anonymization via data aggregation, encryption, etc.~\cite{Ericsson}.\vspace{1em}\\

\hline

\end{tabular}\label{roadmap}}
\caption{A roadmap of big data analytics in DEM}
\end{figure*}

\section{Data Mining and Predictive Analytics in SGs}\label{sec:dataminingsg}
Data mining is the standard process to harvest useful information from a stream of data, such as users' consumption, injection of renewable power, and EVs' state of battery, and transform it into an understandable structure for further use. The data mining process involves the utilization of algorithms for discovering patterns among the data following a similarity criterion~\cite{Chicco}. Efficient data mining is crucial towards the optimized operation of the SG, since it strongly affects the decision-making of power producers and consumers, and the reliability of the grid.

\subsection{Dimensionality Reduction}\label{subsec:dimensionality}
Smart meters generate large volume of data, and thus acquiring and processing all of them is inefficient -if not prohibitive- in terms of communication cost, computing complexity, and data storage resources utilization. For this purpose, dimensionality reduction has been applied in~\cite{Dieb}, in order to provide a reduced version (sketch) of meters' original data via random projection (RP). It is shown that processing the produced summarized version of data instead of the original stream of data leads to an acceptable relative error. The main advantage of RP is scalability, complexity reduction, and execution speed increase.

Dimensionality reduction has only been sufficiently explored in the area of synchrophasor data. Particularly, online dimensionality reduction has been proposed in~\cite{Dahal}, in order to extract correlations between synchrophasor measurements, such as voltage, current, frequency etc. The proposed method can be used a preprocessing method in data analysis and storage, when a only an approximation of the initial data is required. Online dimensionality reduction has also been successfully used in for early event detection in~\cite{LeXie}, where an early event detection algorithm is proposed.

\subsection{Load Classification}\label{subsec:load}
In classification problems a set of pre-classified data points are given and the classification algorithm tries to discover a rule, which describes as closely as possible the observed classification~\cite{Vale}. LC is based on clustering process, which is used to discover groups and identify distributions in the provided data.

For the successive LC in SGs, the most widely used models are Artificial Neural Networks (ANNs). ANNs are computational models consisting of a large number of simple interconnected processors, which can be used to estimate approximate functions that depend on a large number of inputs when there is not an accurate mathematical model to describe the phenomenon~\cite{Zomaya4}. This can be achieved by weighting and transforming the input values by a suitable function with the aid of sequential sets of neurons (until an output neuron is activated). In~\cite{Macedo}, ANNs have been used for the successful classification of consumer load curves in order to create patterns of consumption and facilitate the selection of the appropriate DSM technique. Self-organizing mapping, also known as Kohonen neural network, which is an unsupervised neural networks method, has also been widely used for LC~\cite{Zhou, Verdu}.

Other commonly used algorithms are K-means, which is based on the Euclidean distance between objects, Fuzzy c-means, which is a local search fuzzy clustering method, and hierarchical clustering method, which is a model that can be viewed as a dendrogram~\cite{Zhou}. Due to the ever-growing nature of the SG, a scalable approach is needed for the effective data harvesting and utilization. For this purpose, an effective online clustering has been proposed in~\cite{Monti}, based on unsupervised learning techniques, improving for this case the eXtended Classifier System for clustering (XCSc). This XCSc-based method fits well the dynamic nature of SGs, while it outperforms the offline strategies in terms of the storage system performance~\cite{Asensio}.

\subsection{Short-term Forecasting}\label{subsec:forecasting}
Short-term load forecasting (STLF) has been widely used over the last 30 years and several models have been developed, which can be summarized as follows: i) regression models, ii) linear time-series-based, iii) state-space models, and iv) nonlinear time-series modeling~\cite{Kyriakides}. Moreover, very little progress has been made in the field of the very-short-term (VST) and ultra-short-term (UST) load forecasting in SGs, which, among others, are necessary for the successful self-healing of SG, since they are appropriate for a few minutes’ load forecasting~\cite{Kyriakides, Guan, Han1}. Among the appropriate methods for VST and UST, only basic ANNs have been widely tested. Generally, most of the research has been focused on large aggregated load data, where most individual variations are averaged out by the effect of the law of large numbers. These methods were seldom tried on individual meters or at meter aggregate levels, such as distribution feeders and substation~\cite{Mirowski}. Additionally, as shown in~\cite{Mirowski}, their performance degrades when the number of the considered meters goes down. For this purpose, a short-term load forecasting based on Empirical Mode Decomposition, Extended Kalman Filter and Extreme Learning with Kernel is proposed in~\cite{Tang}, which is more appropriate for load forecasting in micro-grids. Kernel methods has attracted the research interest in the area of STLF~\cite{Mori, Kramer}, since it increases the computing energy-efficiency~\cite{Yoo, Yoo4, Yoo5, Yoo7}. Most works in the existing literature ignore the interdependency between the demand and the electricity prices, while assuming that pricing setting follows the load forecasting. To this end, the authors in~\cite{Motamedi} propose a multi-input multi-output forecasting engine for joint price and demand prediction using data association mining algorithms. For the effective price forecasting, the users' comfort has also to be taken into account, which can be achieved via supervised machine algorithms~\cite{Bei}. For very short term wind power generation forecast, there are myriad of approaches, which can be divided into two big categories: i) forecasting the wind and direction for a specific windmill farm, and ii) forecasting the generated power in a single step~\cite{Courceiro}. More interesting details on this issue can be found in~\cite{Courceiro}. Finally, short-term photovoltaic power prediction is mainly based on the past power output~\cite{Murakami}.

\subsection{Distributed Data Mining}\label{subsec:datamining}
The traditional centralized frameworks for acquiring, analyzing and processing data, require huge exchange of information among the remote sensors (e.g., the smart meters and the centralized processor), which is inefficient in terms of telecommunication resources management and economic cost. To this end, the authors in~\cite{Mallik} present several distributed data analysis techniques that can be successively used for energy demand prediction. The provided analysis emphasizes on the problem of multivariate regression and rank ordering in a distributed scenario, based on polynomially bounded computations per node. Decentralized data mining algorithms have the advantage of scalability, while they are less affected by peer failures and they need little computing and communication resources~\cite{Bhaduri}.

\section{High Performance Computing}\label{sec:HPC}
Real-time monitoring, DEM, and power flow optimization are all based on fast data processing and BDA, which need high computing power. Efficient data mining algorithms based on task parallelism, using multi-core, cluster, and grid computing, can reduce the computational time~\cite{Green}. However, covering the increased data storage and computing resources needs is still a big economic challenge, mainly for the operators of the electricity grids. Therefore, distributed computing seems to be a promising perspective~\cite{Zomaya3}.

\subsection{Dedicated Computational Grid}\label{subsec:grid}
In order to enhance the existing computational capabilities and increase efficiency, a dedicated grid computing based framework is proposed in~\cite{Ali}. In more detail, an architecture of three layers is proposed, namely i) the resource layer, which consists of the hardware part of the computing grid, ii) the grid middleware, which provides access of grid resources to the grid services, and iii) the application layer, which consists of the services. It is shown that this computational grid can provide HPC by combining the processing power, memory and storage of the available computers.

\subsection{Cloud Data}\label{subsec:cloud}
The cloud computing (CC) model meets the requirements of data and computing intensive SG applications~\cite{Rusitschka, Bera}. The main advantages of CC over traditional models are energy saving, cost saving, agility, scalability, and flexibility, since computational resources are used on demand~\cite{Hayes}. Many approaches have been developed so far to further increase the energy efficiency of HPC data centers, such as energy conscious scheduling in~\cite{Zomaya2}, the cooperation with the SG in~\cite{Ghamkhari} and thermal-aware task scheduling in~\cite{Tang}. In~\cite{Rusitschka}, a model for SG data management is presented, taking advantage of the main characteristics of CC computing, such as distributed data management, parallelization, fast retrieval of information, accessibility, interoperability and extensibility. Most of smart grid applications, such as advanced metering infrastructure, SCADA, and energy management, can be facilitated by the available cloud service models, namely software as a service, platform as a service, and infrastructure as a service~\cite{Markovic}.

\subsubsection{Security}\label{subsubsec:security}
Confidentiality and privacy are big challenges towards the application of CC on SG data processing~\cite{Yigit,raey}. To this end, the designed data architectures must be multi-tenant, following one of the three different approaches for such architectures, namely the separate databases, separate schemas, or shared schemas~\cite{Rusitschka}. Also, privacy of end users and data anonymization can be guaranteed by data aggregation, which is used in most SG architectures. However, from the electrical companies' perspective, security is still a challenging problem, since the hackers of systems located in the cloud cannot be easily traced. Authentication, encryption, trust management, and intrusion detection are important security mechanisms that can prevent, detect and mitigate such network attacks~\cite{Ericsson, Ancillotti}. Finally, a related major issue is the recovery of data in the case of a possible failure of the cloud service~\cite{Markovic}.

\section{Future Research Directions \& Discussion}\label{sec:futuredirs}
The load data in SG environment is massive, dynamic, high-di\-mensio\-nal, and heterogeneous~\cite{Zhou}.
Thus, in order to build an accurate real-time monitoring and forecasting system, two novel concepts have to be taken into account in the system design. First, as shown in Fig.~\ref{forecast_model}, all available information from different sources, such as individual smart meters, energy consumption schedulers, aggregators, solar radiation sensors, wind-speed meters and relays has to be integrated, while a communication point has to be designed where multiple artificial experts can interact and make decisions on data. The appropriate forecasting system should rely on effective data sampling~\cite{Yoo3}, improved categorization of the information and successful recognition of the different patterns. Second, suitable adaptive algorithms and profiles for effective dynamic, autonomous, distributed, self-organized and fast multi-node decision-making have to be designed. It has been shown that the performance of multi-node load forecasting is clearly better than that of single-node forecasting~\cite{Han}. The designed algorithms should be based on realistic consensus functions or scoring/voting models~\cite{Hajdu}, where the large computations can be parallelized. The algorithmic results are the state estimation, the estimated production and consumption, and the STLF in SGs. For the most efficient pattern-recognition and state estimation in the SGs environment, the following methodologies and technologies can be used.

\tikzstyle{cloud} = [draw, ellipse,fill=green!5, minimum height=3em]
\tikzstyle{block} = [rectangle, draw, fill=blue!5, text centered, rounded corners, minimum height=0em]
\tikzstyle{block1} = [text centered, minimum height=0em]
\tikzstyle{block2} = [rectangle, draw, fill=green!5, text centered, rounded corners, minimum height=2.9em]
\tikzstyle{block3} = [rectangle, fill=green!5, text centered, rounded corners, minimum height=0em]
\tikzstyle{block4} = [rectangle, draw, fill=blue!5, text centered, rounded corners, minimum height=19.7em, minimum width=24em]
\tikzstyle{block5} = [text centered, draw, dashed, fill=blue!5, minimum height=0em]
\tikzstyle{line} = [draw, thick, -latex']

\begin{figure*}
\centering%
\resizebox{1\linewidth}{!}{%
\begin{tikzpicture}
    \node [cloud] (data) {Data Sources};
    \node [block2, right=0.8cm of data] (extraction) {Feature Extraction};
    \node [block, below=0.8cm of extraction] (extraction1) {\begin{tikzpicture} \node [block1] (imaginary) {}; \node [block1, left=2.2cm of imaginary] (imaginary1) {}; \node [block1, right=2.1cm of imaginary] (imaginary2) {}; \node [block3, below= 0cm of imaginary1] (trad_factors) {\begin{tabular}{|l|c|}
\hline \textbf{Traditional Factors}\\
\hline \hline
Weather\\
\hline
Time of the Day \\
\hline
Day of the year\\
\hline
Random Events\\
\& Disturbances\\
\hline

\end{tabular}};
    \node [block3, below= 0cm of imaginary2] (smartfactors) {\begin{tabular}{|l|c|}
\hline \textbf{Smart Grid Factors}\\
\hline \hline
Electricity Prices\\
\hline
Demand Response\\
\hline
Distributed Energy Resources\\
\hline
Storage Cells\\
\hline
Electric Vehicle\\
\hline
\end{tabular}}; \end{tikzpicture}  };

    \node [block4, right= 0.8cm of extraction1] (keno){};
    \node [block1, right= 4.4cm of extraction] (map) {\textbf{Map Reduce Parallel Computing Platform}};
    \node [block2, below= 0.4cm of map] (pre) {Preprocessing, i.e., dimensionality reduction};
    \node [block2, below= 0.5cm of pre] (build) {Building $m$ models (Training) for previous $n$ Days};
     \node [block2, below=0.5cm of build] (real) {Real-Time Load Forecasting};
     \node [block2, below=0.5cm of real] (post) {Post Processing};
     \path [line] (pre) -- (build);
\path [line] (build) -- (real);
\path [line] (real) -- (post);
    \node [block2, below= 0.8 of keno] (load) {Load Forecast Value in STLF, VST, UST};
 \node [block5, right= 3cm of load] (update) {Update};
   \node[block1, left=1 of load](caption_imaginary){};
   \node[block1, below=0.6 of caption_imaginary](caption_imaginary){\textbf{Proposed Smart Grids forecast model}};
    \path [line] (data) -- (extraction);
    \path [line] (extraction) -- (extraction1);
    \path [line] (extraction1) -- (keno);
     \path [line] (keno) -- (load);
    \path [line,dashed] (update)|- (build);
\path [line,dashed] (load)|- (update);
\end{tikzpicture}}
\caption{Proposed SG forecast model.}\label{forecast_model}
\end{figure*}

\subsection{SG Feature Selection and Extraction}\label{subsec:sgselection}
The factors that affect the load forecasting can be separated in two categories: a) the traditional factors and b) the SG factors~\cite{Balantrapu}. The traditional factors include the weather conditions, time of the day, season of the year, and random events and disturbances. On the other hand, the smart grid factors include the electricity prices, demand response, distributed energy sources, storage cells and electric vehicles. As shown in Fig.~\ref{forecast_model}, large volumes of data from the sensors installed around SG are collected, and the features extracted in this phase also need to be refined as there is noise and redundancy in the features. If the input features contain redundant information (e.g., highly correlated features), ML algorithms in general perform poorly because of numerical instabilities. Some regularization techniques can be imposed to solve such problems. The techniques that can be used to build an optimal subset for the load predicting problem in SG are greedy hill climbing~\cite{Alajmi}, minimum-redundancy-maximum-relevance~\cite{Peng}, regularized trees~\cite{Houtao}, random multinomial logit~\cite{Anita}, etc. In addition to the image preprocessing methods described above, there are many other techniques that draw on disciplines such as artificial intelligence (AI) and ML that can be used to analyze such selected features.

\subsection{Online Learning}\label{subsec:learning}
Online learning is a powerful way of dealing with load monitoring and prediction in SGs. An online learning algorithm observes a stream of examples and makes a prediction for each element in the stream~\cite{Botao}. The algorithm receives immediate feedback about each prediction and uses this feedback to improve its accuracy on subsequent predictions. In contrast to statistical ML, online learning algorithms don't make stochastic assumptions about the observed data, and even handle situations where the data is generated by a malicious adversary. There has been a recent surge of scientific research on online learning algorithms, largely due to their broad applicability to web-scale forecasting problems. In load forecasting in SGs, the statistical ML properties of the target variable, which the model is trying to forecast, change over time in unforeseen ways. This causes the so-called ``concept drift" problem because the predictions become less accurate as time passes. In~\cite{Anderson}, the employed online learning set up mitigates such problem. In online learning, soon after the forecasting is made, the true label of the instance is discovered. This information can then be used to refine the forecasting hypothesis used by the algorithm. The goal of the algorithm is to make forecasting that is close to the true labels.

\subsection{Randomized Model Averaging}\label{subsec:randmodel}
ML concerned with the design and development of algorithms that allow computers to evolve behaviours based on empirical data. A major focus of ML research is to automatically learn to recognize complex patterns such as the features of SGs, and make intelligent decisions based on data. Using multiple predictive ML models (each developed using statistics and/or ML) to obtain better predictive performance than could be obtained from any of the constituent models~\cite{Lauri, Yoo8}. In addition, the proposed scheme should be utilizing the model averaging technique~\cite{Hoeting} in order to improve the stability and accuracy of ML algorithms via reducing variance~\cite{Yoo, Yoo2, Yoo6}.

\subsection{MapReduce Parallel Processing}\label{subsec:mapreduce}
Load forecasting problems, involving intelligent ML solutions pertaining to large volume of data source generation, are a perfect fit for MapReduce deployments~\cite{Botao}. MapReduce is a programming model for processing large datasets with a parallel, distributed algorithm on a computing cluster of low cost commodity computers. A MapReduce application typically consists of two phases (or operations); ``map'' and ``reduce'' with many tasks in each phase~\cite{Zomaya5}. A map/reduce task deals with a chunk of data independently and thus, tasks in a given phase can be easily parallelized and effectively processed in a large-scale computing environment (i.e., a cloud platform). However, the existing resource management models pay little attention to the data and/or utilize a simple technique, called MapReduce’s rack-aware task placement. Such parallelization model could be reconstructed by explicitly taking the application characteristics and network topology into account~\cite{forest, Yoo9}. For instance, in a tree network topology, two sub-trees (racks) adjacent to each other will be a better combination than two of dispersed sub-trees. The presented framework can enable the minimization of data movement and in turn reduce the occurrence of network contention, and this will eventually enhance the cloud system’s efficiency. Two main mechanisms that have to be taken into account explicitly are a) the data locality-aware scheduling algorithm, and b) the application-specific resource allocation mechanisms. Specifically, tasks requiring common datasets are dispatched to computers (compute nodes) with close proximity to those data sets. For most data processing applications, storage capacity, such as disk and memory, is more important than computing power (i.e., CPUs).

\subsection{Available Testbeds and Platforms}\label{subsec:testbeds}
The majority of the available power grid systems focus on modeling of traditional network components, i.e., the generation systems, loads, and transmission network. A different approach is followed in~\cite{Ning}, where a distribution grid testbed has been proposed, which can be used to test the designs of integrated information management systems. The purpose of this testbed is to successfully represent the correlation and interdependency among data sets, aiming to efficiently monitor the status of the SG and detect abnormalities. Interestingly enough, extensive sets of SG's detailed trial data, which can be used in order to test the designed schemes, can be easily acquired, thus facilitating the research in this area~\cite{data, Hao, data1, data2, data3}. In~\cite{data}, registered users can also access the public model, including key functions, assumptions, and analytical tools.

Storing and processing the huge amount of data generated by the smart meters, requires improved platforms, appropriate for big data analytics, such as Hadoop, Cassandra, and Hive~\cite{Mayilvaganan}. Hadoop is a promising platform for the distributed processing of large SG's data sets. It is s a collection of open source tools and includes the concept of MapReduce. Cassandra database, which supports the cloud infrastructure, can be used in order to store the large data sets that are needed for the effective DEM. Moreover, Hive data warehouse software, which uses a simple SQL-like language, can be used to query datasets that are stored in a distributed environment.

\section{Conclusions}\label{sec:conclusion}
In this paper, we have summarized the state-of-the-art in the exploitation of big data tools for dynamic energy management in smart grid platforms. We have first highlighted that, in order to deal with the extreme size of data, the smart grid requires the adoption of advanced data analytics, big data management, and powerful monitoring techniques. Next, we elaborated on the utilization of the most commonly used smart grid data mining and predictive analytics methods, focusing on the smart meter data that are necessary for the accurate and efficient power conumption/supply forecasting. We proceeded with a brief survey on the works dealing with high performance computing, insisting on cost efficiency and security issues in the context of SG control. Finally, we discussed several interesting techniques and methods that have to be further explored into the framework of a real-time monitoring and forecasting system, and we provided promising research directions for future research in the field.

\section*{Acknowledgement}
This work was supported by the project of ``Scalable Real-Time Load Forecasting in Smart Grid" under a grant from the Khalifa University Internal Research Fund (KUIRF-Level 2; Fund $\#$ 210063).

\bibliographystyle{elsarticle-num}\balance

\end{document}